\documentclass[twocolumn,showpacs,preprintnumbers,amsmath,amssymb]{revtex4}

\usepackage{graphicx}
\usepackage{dcolumn}
\usepackage{bm}

\begin{document}

\preprint{APS/123-QED}

\title{Some aspects of the synchronization in coupled maps}

\author{Sandro E. de Souza Pinto}
\email{sandroesp@uepg.br}
\author{Jos\'{e} T. Lunardi}
\author{Abdala M. Saleh}
\author{Antonio M. Batista}

\affiliation{Grupo de F\'{\i}sica Te\'{o}rica, Departamento de
Matem\'{a}tica e Estat\'{\i}stica, Universidade Estadual de Ponta
Grossa\\Av. Gal. Carlos Cavalcanti 4748. CEP 84032-900, Ponta
Grossa, PR, Brazil}

\date{\today}

\begin{abstract}
Through numerical simulations we analyze the synchronization time
and the Lyapunov dimension of a coupled map lattice consisting of a
chain of chaotic logistic maps exhibiting power law interactions.
From the observed  behaviors we find a lower bound for the size $N$
of the lattice, independent of the range and strength of the
interaction, which imposes a practical lower bound in numerical
simulations for the system to be considered in the thermodynamic
limit. We also observe the existence of a strong correlation between
the averaged synchronization time and the Lyapunov dimension. This
is an interesting result because it allows an analytical estimation of the
synchronization time, which otherwise requires numerical simulations.

\end{abstract}


\maketitle

Coupled Map Lattices (CML's) were introduced in the literature as
suitable models to study spatiotemporal behavior of spatially extended
dynamical systems. Fundamentally they are systems defined on a
discrete space-time and possessing continuous state variables. In
the last two decades such models have received a great and
increasing deal of interest, being applied to several nonlinear
phenomena including systems as diverse as physical, chemical and
biological\cite{kaneko1}. CML's share with real complex systems one
of their most intriguing behavior, which is the possibility of
synchronization. Such a phenomenon can be observed in a great
variety of real systems, going from electronic circuits until
physiological processes, for example \cite{boccaletti, synchro}.

The model we are concerned here consists of a chain of $N$ identical
chaotic logistic maps \cite{livrott}, each one located at a definite
site in a discrete space, and coupled between themselves through a
power law interaction \cite{pintoviana}. Our aim is to go a step
further in the understanding the synchronization of this system
through the analysis of the synchronization time and the Lyapunov
dimension. 

The model is defined as follows. At the (discrete) instant of time $n$ the state, or amplitude, of
the map located at the site $i$ ($i=1,2,\cdots,N$) is denoted by the
continuous variable $x_n^{(i)}$. The state of the whole lattice at
time $n$ will be given by the $N$-dimensional vector
${\bf{x}}_n=\left(x_n^{(1)}, x_n^{(2)},\cdots ,x_n^{(N)}\right)$.
The time evolution of the system is given, in matrix form, by the
following mapping
\begin{equation}
{\bf {x}}_{n+1}={\bf F}({\bf {x}}_n) + {\bf I}\;{\bf F}({\bf
{x}}_n)\, ,\nonumber
\end{equation}
where ${\bf F}({\bf x})=\left(f(x^{(1)}), f(x^{(2)}),\dots,
f(x^{(N)})\right)$, and ${\bf I}$ is an $N\!\times\!N$ matrix which
specifies the coupling among the elemental maps. The function $f(x)$
characterizes the elemental map, which we assume here to be the
fully chaotic logistic one $f(x)=4x(1-x)$, with $x\in [0,1]$. We
always assume odd $N$ and periodic boundary conditions. The initial
state of each elemental map is randomly chosen. The power law
coupling is given by
$$
I_{ij}=\frac{\varepsilon}{\eta}\left\{\frac{1}{r_{ij}^\alpha}
\left(1-\delta_{ij}\right)-\eta\delta_{ij}\right\}\, ,
$$
where $I_{ij}$ are the matrix elements of ${\bf I}$ and
$r_{ij}=|i-j|$ is the ``distance" between sites $i$ and $j$. The
parameters $\varepsilon$ and $\alpha$ give, respectively, the
strength  and the range of the interaction and
$\eta=2\sum_r^{N'}r^{-\alpha}$ is a normalization factor, with
$N'=(N-1)/2$. The parameter $\alpha$ can assume any value in the
interval $[0,\infty)$. The extreme cases $\alpha=0$ and
$\alpha\to\infty$ correspond, respectively, to \emph{global} (mean
field) and \emph{local} (first neighbors) couplings. Here we
restrict the parameter $\varepsilon$ to the interval $[0,1]$ in
order to get all the individual map amplitudes into the interval
$[0,1]$. In a recent work, including one of us, it was
proposed that the parameter $\varepsilon$ could assume any
nonnegative value \cite{celia2}. Although both systems exhibit the
same behavior when $0\leq\varepsilon\leq 1$, our choice implies that
the nonlinearity of the model arises \emph{solely} from the
nonlinearity of the elemental maps, while the choice in ref.
\cite{celia2} allows, for $\varepsilon>1$, also nonlinearities
arising from the $\texttt{mod}\, 1$ operation in the coupling
scheme.

Among the various characterizations of synchronization
\cite{boccaletti} we choose the following. A system is said to be in
a \emph{completely synchronized state} at time $n$ if all the
elemental maps have the same amplitude, i.e.,
$x_n^{(1)}\!=\!x_n^{(2)}\!=\!\cdots\!=\!x_n^{(N)}\!=\!x_n^{*}$. The
subspace ${\bf S}$ of all these states (the diagonal of the whole
state space of the system) will be called \emph{invariant} if
${\bf{x}}_{m}\!\in\! {\bf S}$ implies that ${\bf{x}}_n\!\in\! {\bf
S}$ for all $n\!>\!m$. In this case a \emph{synchronized regime}
starts when the system state is put on ${\bf S}$, which is then
called the \emph{synchronization subspace}. Given specific initial
conditions, the minimum value of $m$ for which the system goes into
this subspace is identified as the \emph{synchronization time}
$t_{\!s}$. The necessary and sufficient condition for ${\bf S}$ to
be  invariant is that the sum $\sum_{j} I_{ij}$ be
independent of $i$, i.e., the sum of all elements of each line of
the matrix $\textbf{I}$ must be the same for all lines
\cite{pecora}. Besides that, in order to hold $x_{n+1}^{(i)}$ in the
interval $[0,1]$, for all $n$ and $i$, it is necessary that this sum
be restricted to the interval $[-1,0]$. In our case $\sum_j
I_{ij}=0$ for all $i$ and ${\bf S}$ turns out to be an invariant
subspace.

\begin{figure}[htb]
\includegraphics[height=5.0cm,width=0.8\columnwidth,clip]{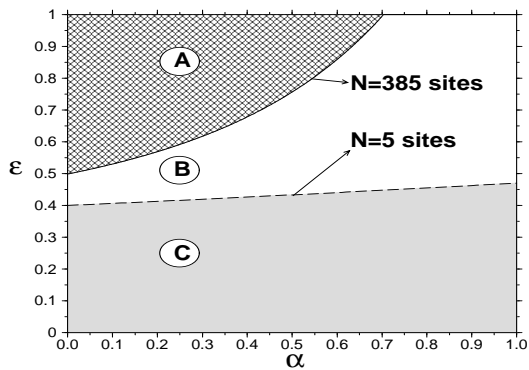}
\caption{\label{figure1} Domains of synchronization in the
parameters space. For $N\!=\!5$ the CML synchronizes in Regions (A)
and (B) and does not in Region (C). For $N\!=\!385$ it synchronizes
in Region (A) and does not in Regions (B) and (C).} 
\end{figure}

If initially the lattice is not synchronized, it only will
synchronize after some time $m$ if the parameters $\alpha$ and
$\varepsilon$ assume values into a suitable domain, for each $N$, in
the parameter space \cite{validity, celia1}. The boundaries of this
domain can be calculated analytically from the condition
$\lambda_2=0$, where $\lambda_2$ is the second largest Lyapunov
exponent of the system, which is the largest Lyapunov exponent
transversal to the synchronization subspace \cite{gade}. These
boundaries can be decomposed into an upper line, given by
$\varepsilon_c^{'}(\alpha,N)\!=\!3/2(1-b^{(N')}/\eta)^{-1}$, and
into a lower one, given by $\varepsilon_c(\alpha,N)\! =\!
1/2(1-b^{(1)}/\eta)^{-1}$, where
$b^{(k)}\!=\!2\sum_{m=1}^{N'}\cos(2\pi km/N)m^{-\alpha}$ $(1\!\leq\!
k\! \leq\! N)$ are the eigenvalues of the circulant matrix ${\bf
B}\!=\! (\eta\varepsilon^{\!-\!1}{\bf I}\! -\!
\eta\varepsilon\openone)$ \cite{celia2, celia1}. For instance, the
domains for $N\!=\!5$ and $N\!=\!385$ are shown in Figure
\ref{figure1}. The shown lines are only the lower ones, for both
$N$; the upper lines do not appear because in the range of figure
all values of $\varepsilon_c^{'}(\alpha,N)$ are greater than $1.0$,
which are outside the range of this parameter. We can observe two
distinct regions for each $N$. For $N\!=\!385$, and if it does not
start synchronized, the lattice synchronizes after some time only in
Region (A), while for $N\!=\!5$ it synchronizes only in Regions (A)
and (B).

On the state space of the system it can be defined the quantity
$d_n\!=\!\sigma_n\sqrt{N}$, where $\sigma_n$ is the standard
deviation of the $x_n^{(i)}$ around the mean $\langle x_n^{(i)}
\rangle$, with the averages taken with respect to the index $i$. It
turns out that this quantity corresponds to the distance, in the
state space, from the point $\textbf{x}_n$ to the synchronization
subspace ${\bf S}$ \cite{distancia}. We thus use the condition
$d_n\!=\!0$ as a diagnostic for the synchronization regime.
Accordingly, the synchronization time $t_s$ is identified with the
lowest $n$ for which $d_n\!=\!0$ \footnote{Due to the limited
precision of numerical simulations, $d$ is considered equals to zero
if $d\!<\!10^{-q}$, where $q$ is the precision, which in our
simulations is set to $16$ (double precision).}.
\begin{figure}[htb]
\includegraphics[width=0.8\columnwidth,clip]{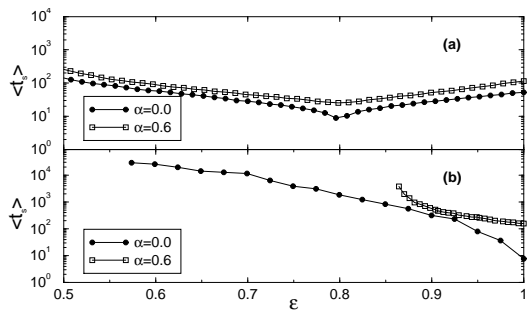}
\caption{\label{figure2} Averaged synchronization time $<\!t_s\!>$
{\it versus} $\varepsilon$, with $\alpha\!=\!0.0$ and
$\alpha\!=\!0.6$ . (a) N\!=\!5; (b) N\!=\!385.}
\end{figure}

Figure \ref{figure2} is a plot of the averaged synchronization time
$\left<t_s\right>$  \emph{versus} the strength parameter
$\varepsilon$, for two values of $N$ and $\alpha$ \footnote{The
average is taken with respect to a sample of 50 randomly chosen
initial conditions.}.  For $N\!=\!5$ (Figure \ref{figure2}(a)), and
for both the values of $\alpha$, we observe an atypical behavior for
the synchronization time when $\varepsilon\!>\! 0.8$. It would be
expected that $\left<t_s\right>$ were always a decreasing function
of the interaction strength, because greater values of this
parameter would tend to facilitate synchronization. We have
constructed similar plots for various values of $N$ and $\alpha$ and
our results indicate that this behavior is characteristic for small
lattices, being therefore a finite size effect of the system. Such a
behavior was not observed with respect to the $\alpha$ parameter.
From our simulations we observed that the ``turning point"
$\varepsilon\!=\!\varepsilon^{(1)}$, above which $\left<t_s\right>$
starts to increase, depends only on $N$ (of course, with $\alpha$
inside the synchronization domain) and it can be clearly identified
as
\begin{equation}
\varepsilon^{(1)} =
\frac{\varepsilon'_c(0,N)-\varepsilon_c(0,N)}{2}.
\end{equation}
It turns out that the values of $N$ for which this atypical behavior
disappears must correspond to $\varepsilon^{(1)}\!\geq\!1$. From the
above expression (or from the numerical simulations) we can observe
that this will be satisfied for $N\!\geq\!N_{min}\!=\!385$. The
behavior of $\left<t_s\right>$ for $N\!=\!N_{min}$ is shown in
Figure \ref{figure2}(b).
\begin{figure}[htb]
\includegraphics[width=0.8\columnwidth,clip]{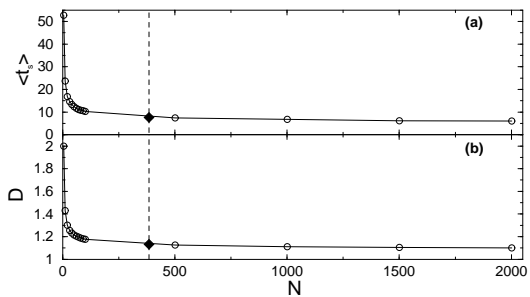}
\caption{\label{figure3} (a) $\left<t_s\right>$ \emph{versus} $N$,
for $\alpha\!=\!0.0$ and $\varepsilon\!=\!1.0$. (b) Lyapunov
dimension $D$ \emph{versus} $N$, for the same values of $\alpha$ and
$\varepsilon$. The vertical line corresponds to $N\!=\!385$.}
\end{figure}

In Figure \ref{figure3} $(a)$ we plot the averaged synchronization
time with varying $N$, with both $\varepsilon$ and $\alpha$ fixed.
The results indicate that $\left<t_s\right>$ tends to saturate for
large $N$. Moreover, the saturated time does not differs
significantly from that corresponding to $N\!=\!385$.

\begin{figure}[htb]
\includegraphics[height=5cm, width=0.8\columnwidth,clip]{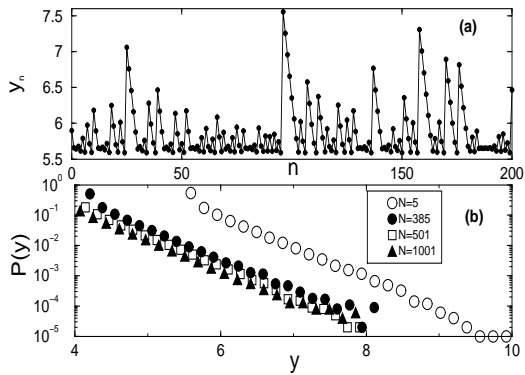}
\caption{\label{figure4} (a) Time series for $y_n$ with $N\!=\!385$,
$\alpha=0.6+10^{-5}$ and $\varepsilon=\varepsilon_c(0.6,1001)$. (b)
Distributions of $y$ for $N\!=\!5$ (open circles), $N\!=\!385$
(filled circles), $N\!=\!501$ (open squares), and $N\!=\!1001$
(filled triangles), with $\alpha=0.6+10^{-5}$ and
$\varepsilon=\varepsilon_c(0.6,N)$.}
\end{figure}

Now we consider the CML in an outer vicinity of the synchronization
domain in parameter space. The system does not more attain the
synchronized regime if it not started synchronized. Figure
\ref{figure4}(a) is a typical plot for a time series of the
distances $d_n$ in such a case. To make easier the visualization we
presented our results in terms of $y_n\!=\!-\log_{10} d_n$. Figure
\ref{figure4}(b) shows the corresponding statistical distributions
of $y_n$. Again the results suggest that, for large lattices, such a
distribution is practically independent of $N$ and it is very well
approximated by the distribution corresponding to $N\!=\!385$.

All the results presented so far indicate that $N\!=\!385$, apart
from eliminating the just mentioned finite size qualitative effects,
also lead to a reasonable quantitative approximation to the behavior
of the system for larger $N$ values, at least in what concern the
synchronization time behavior and the statistical distribution of
the distances $d_n$. So, at least in what concerns these two
aspects, we claim that $N_{\emph{min}}$ sets a practical lower bound
in numerical simulations for the system to be considered at the
thermodynamic limit. With this claim we mean that both the
qualitative and quantitative behaviors of the system at the
thermodynamic limit $N\!\to\!\infty$ can be reasonably well
approximated by the corresponding behavior of the system with
$N\!=\!N_{\emph{min}}$.

\begin{figure}[htb]
\includegraphics[height=5cm,width=0.8\columnwidth,clip]{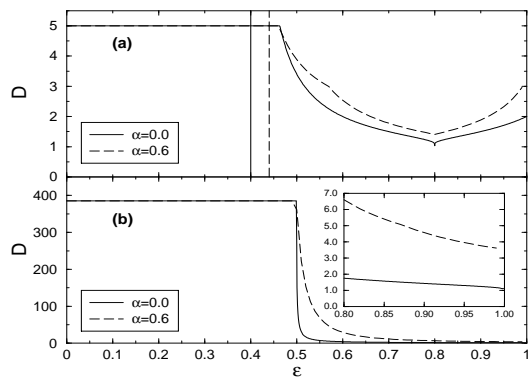}
\caption{\label{figure5} Lyapunov dimension $D$ {\it versus}
$\varepsilon$ with $\alpha=0.0$ and $\alpha=0.6$. (a) $N=5$; (b)
$N=385$. The vertical lines indicate the boundary of the
synchronization domain. \emph{Inset}: detail of (b) for $0.8\leq
\varepsilon\leq 1.0$}
\end{figure}

Now, we recall that the time oscillation of $d_n$ in Figure
\ref{figure4}(a) is due to the coexistence of both stable and
unstable Lyapunov exponents in the direction transversal to the
invariant subspace {\bf S} \cite{validity, celso}. Therefore, a
closer examination of the Lyapunov spectrum could reveal some new
aspects of the synchronization behavior.  We thus consider the
Lyapunov dimension of the system, which is a suitable concept to
study the Lyapunov spectrum and is defined as follows. Let
$\lambda_j$ $(j=1,2,\cdots)$ denote the $j$-th largest Lyapunov
exponent of the system and $p$ be the largest integer for which
$\sum_{j=1}^p\lambda_j$ is non negative. Then $D$ is given by
\cite{livrott}
\begin{equation}
D=\left\{\begin{array}{ll} 0 &
\mbox{if there is no such \textit{p}} \\
p+\frac{1}{|\lambda_{p+1}|}\sum_{i=1}^p\lambda_i & \mbox{if $p<N$}\\
N &\mbox{if $p=N$} \end{array}\right.
\end{equation}
In Figure \ref{figure5} we depict the Lyapunov dimension $D$
\emph{versus} the strength parameter $\varepsilon$, for $N\!=\!5$
and $N\!=\!385$, and for two values of $\alpha$. We can observe that
$D$ assumes a maximum $D_{max}\!=\!N$ for small values of
$\varepsilon$. As $\varepsilon$ enters into the synchronization
domain, identified in the figure by vertical lines, we can observe
two distinct behaviors. For large $N$, it monotonically decreases,
but for small $N$ there is a value
$\varepsilon\!=\!\varepsilon^{(2)}$ above which $D$ starts to
increase.
Figure \ref{figure3} $(b)$ shows the dependence of $D$ with $N$, for
$\alpha$ and $\epsilon$ fixed within the synchronization domain. We
can observe that the Lyapunov dimension tends to saturate with
increasing $N$.

\begin{figure}[htb]
\includegraphics[height=5cm,width=0.8\columnwidth,clip]{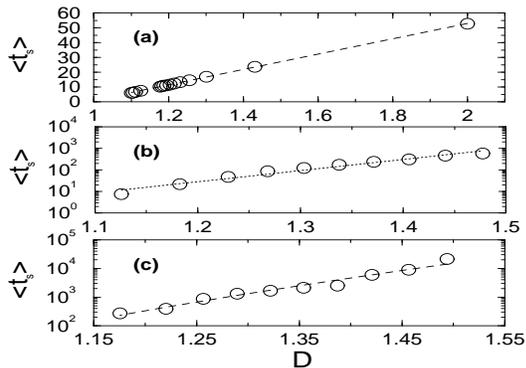}
\caption{\label{figure6} Dispersion diagrams for
$\left<t_s\right>\times D$: (\emph{a}) $\alpha\!=\!0.0$,
$\varepsilon\!=\!1.0$ and $N$ ranging from $N\!=\!5$ to
$N\!=\!2000$, $\rho\!=\!0.9998714$; (\emph{b}) $\alpha\!=\!0.0$,
$N\!=\!501$ and $\varepsilon$ ranging from
$\varepsilon\!=\!\varepsilon_c(0,501)$ to $\varepsilon\!=\!1.0$,
$\rho\!=\!0.9849924$; (\emph{c}) $N\!=\!501$, $\varepsilon\!=\!1.0$
and $\alpha$ ranging from $\alpha\!=\!0.0$ to $\alpha\!=\!0.2$,
$\rho\!=\!0.9868354$.}
\end{figure}

We point now the great similarity between the behavior of the
Lyapunov dimension within the synchronization domain and the
behavior observed for the averaged synchronization time
$\left<t_s\right>$, as can be seen by a direct comparison between
the shapes of figures \ref{figure2}  and \ref{figure5} or between
figures \ref{figure3} $(a)$ and $(b)$. These plots suggests that
there is a correlation among the Lyapunov dimension and the averaged
synchronization time. In Figure \ref{figure6} we plot three
dispersion diagrams $\left<t_s\right>\times D$, each one with two of
the three parameters $\alpha$, $\varepsilon$, $N$ fixed. All these
diagrams give correlation coefficients $\rho$ very close to $1$,
which indicates a \emph{very strong} correlation among these two
quantities. In these plots the dashed lines correspond to the
fitting functions. In the first case the fitting is linear and in
the two last they are exponentials. As the Lyapunov dimension can be
analytically determined, this result allows also an analytical
estimation of the synchronization time, which otherwise requires
numerical simulations.  The origin of such strong
correlation between so diversely defined quantities is a point that
would need a deeper analysis, which is out the scope of this report.
Nevertheless, we could try to understand this fact on some intuitive
grounds by  observing that it is
reasonable to think that the dominance of negative(positive)
Lyapunov exponents in the direction transversal to the invariant
subspace ${\bf S}$ would tend to minimize(maximize) the
synchronization time. Accordingly, as it is straightforward from its
definition, this is precisely the behavior of the Lyapunov
dimension.

Summarizing, in this report we numerically simulated the behavior of
a CML consisting of a chain of chaotic logistic maps exhibiting
power law interactions. We observed size dependent behaviors with
respect to the averaged synchronization time $\left<t_s\right>$ and
with respect to the statistical distribution of the distances $d_n$,
which allowed us to set $N\!=\!385$ as a practical lower bound for
this system to be considered in the thermodynamic limit in numerical
simulations. We argued that the system behavior at the thermodynamic
limit can be reasonably well approximated, both qualitatively and
quantitatively, by its behavior at this lower bound. By the way, the
behavior of systems exhibiting long range couplings, specially
concerning their thermodynamical aspects, is still not well
understood. We hope that these results could give some useful
contribution to this subject.

The Lyapunov dimension for the system within the synchronization
domain was studied and the results showed a very strong correlation
among it and the averaged synchronization time. This result seems
interesting  because it allows an analytical estimation of the
synchronization time, which otherwise requires numerical simulations. 
The origin of such a correlation and its related
consequences are subjects that still need more clarifications and it
will be postponed to future works. Finally, we remark that the
statistical distribution of distances in Figure \ref{figure2} can be
well fitted by a power law. Further studies on this and other
related scaling laws on this system are now in curse and shall be
presented elsewhere.

\begin{acknowledgments}
The authors thank the supports from CNPq (SESP and AMB) and
CNPq/Funda\c{c}\~ao Arauc\'aria (JTL). SESP and AMB also thank
Ricardo L. Viana for fruitful discussions.
\end{acknowledgments}

\end{document}